\documentclass[journal]{IEEEtran}
\usepackage{color}

\ifCLASSINFOpdf
   \usepackage[pdftex]{graphicx}
   \DeclareGraphicsExtensions{.pdf,.jpeg,.png}
\else
\fi
\ifCLASSOPTIONcompsoc
 \usepackage[caption=false,font=normalsize,labelfont=sf,textfont=sf]{subfig}
\else
 \usepackage[caption=false,font=footnotesize]{subfig}
\fi
\usepackage{url}

\usepackage{multirow}
\usepackage{hyperref}

\begin{document}
%
\title{Txt2Vid: Ultra-Low Bitrate Compression of Talking-Head Videos via Text}
%
%
%


\author{Pulkit Tandon,
        Shubham Chandak,
        Pat Pataranutaporn,
        Yimeng Liu,
        Anesu M.\ Mapuranga,\\
        Pattie Maes,
        Tsachy Weissman,
        and~Misha Sra
\thanks{P. Tandon, S. Chandak, A.M. Mapuranga and T. Weissman are with the Department
of Electrical Engineering, Stanford University, Stanford, CA, 94305 USA.}
\thanks{P. Pataranutaporn and P. Maes are with the MIT Media Lab, MIT, Cambridge, MA, 02139 USA.}
\thanks{Y. Liu and M. Sra are with the Department of Computer Science, UC Santa Barbara, Santa Barbara, CA, 93107 USA}
\thanks{Corresponding Author e-mail: tpulkit@stanford.edu}}

%
%

\markboth{Journal of \LaTeX\ Class, 2021}%
{Tandon \MakeLowercase{\textit{et al.}}: Txt2Vid: Ultra-Low Bitrate Compression of Talking-Head Videos via Text}
%



\maketitle

\begin{abstract}
Video represents the majority of internet traffic today, driving a continual race between the generation of higher quality content, transmission of larger file sizes, and the development of network infrastructure. In addition,  the recent COVID-19 pandemic fueled a surge in the use of video conferencing tools. Since videos take up considerable bandwidth ($\sim$$100$ Kbps to a few Mbps), improved video compression can have a substantial impact on network performance for live and pre-recorded content, providing broader access to multimedia content worldwide. We present a novel video compression pipeline, called Txt2Vid, which dramatically reduces data transmission rates by compressing webcam videos (``talking-head videos'') to a text transcript. The text is transmitted and decoded into a realistic reconstruction of the original video using recent advances in deep learning based voice cloning and lip syncing models. Our generative pipeline achieves two to three orders of magnitude reduction in the bitrate as compared to the standard audio-video codecs (encoders-decoders), while maintaining equivalent Quality-of-Experience based on a subjective evaluation by users ($n=242$) in an online study. The Txt2Vid framework opens up the potential for creating novel applications such as enabling audio-video communication during poor internet connectivity, or in remote terrains with limited bandwidth. The code for this work is available at \url{https://github.com/tpulkit/txt2vid.git}.
\end{abstract}


%
\IEEEpeerreviewmaketitle

\section{Introduction}
\label{sec_intro}

\begin{figure*}[!t]
\centering
\includegraphics[width=\textwidth]{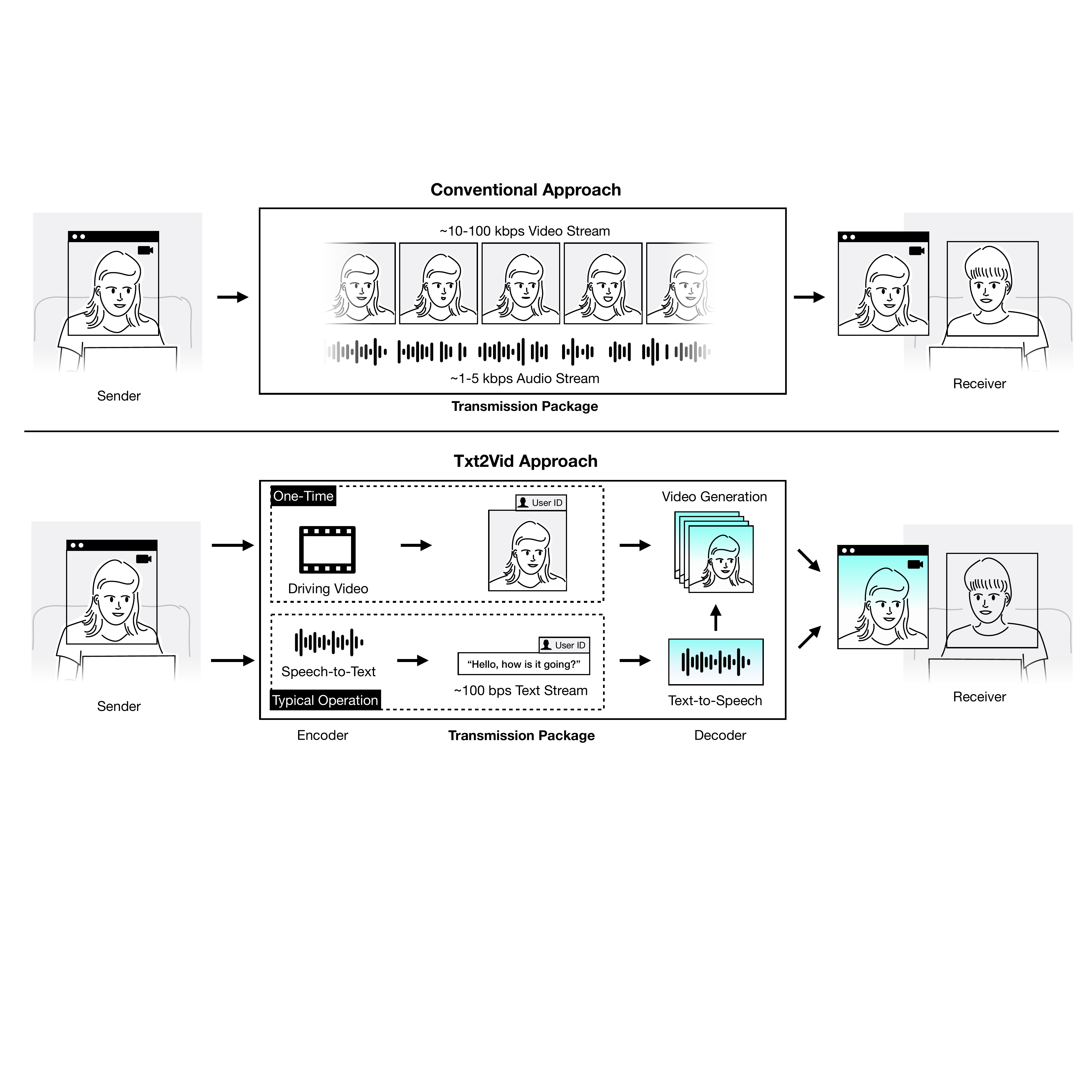}
\caption{Motivation. In traditional video codecs (encoders-decoders) an audio-video stream of the sender is transmitted to the receiver which requires a bandwidth of $\sim$$10$-$100$ Kbps. On the other hand, in Txt2Vid, videos are compressed to text which can be transmitted at a bitrate of only $\sim$$100$ bps and recovered at the decoder using state-of-the-art deep learning based models.}
\label{fig_motivation}
\end{figure*}



Video streaming represents the majority share of internet traffic today, with estimates as high as $80\%$ \cite{cisco2017}. With the COVID-19 outbreak, internet services have seen a surge in usage ($\sim$$50$-$100\%$), with video conferencing tools such as Zoom seeing ten times increase in usage \cite{pandey2020impact}. A typical video conferencing call can consume anywhere from $\sim$$100$ Kbps to a few Mbps (bps refers to bits-per-second). Unfortunately, a vast majority of the world's population does not have access to high bandwidth network connections \cite{cisco2018} or faces intermittent connectivity issues. The ability to conduct an audio-video (AV) call at extremely low bitrates ($\sim$$100$-$1000$ bps) could provide broader access to billions of people in developing countries or other locations with limited or unreliable broadband connectivity. Even with major improvements in bandwidth, advances in compression of data generated from video conferencing tools can provide broader access of this technology worldwide. Moreover, a reduction in required bandwidth can have a significant impact on global network performance by decreasing the network load \cite{candela2020impact, ford2020covid}. Thus, efficient compression of data generated using video conferencing tools is an important problem.

In this work, we propose and demonstrate a pipeline which can compress videos to text and reconstruct the videos from text using state-of-the-art deep learning based decoders. Given the recent extreme need and reliance on video conferencing, the focus of our work is on audio-video (AV) content transmitted from webcams during video conferencing or webinars. Current compression codecs (such as H.264 \cite{wiegand2003overview} or AV1 \cite{chen2018overview} for videos and AAC \cite{brandenburg1999mp3} for audio) lossily compress the input AV content by discarding details that have the least impact on user experience. However, the distortion measures targeted by these codecs are often low-level and attempt to penalize deviation from the original pixel values or audio samples. But in reality, what matters most is the final quality-of-experience (QoE) when a compressed media stream is shown to a human end-consumer \cite{chen2014qos}. Thus, in our proposed pipeline, instead of working with pixel-wise fidelity metrics we recreate the original content such that the QoE is maintained. Figure \ref{fig_motivation} contrasts the conventional compression approaches with our Txt2Vid approach. While conventional approaches typically require $\sim$$10$-$100$ Kbps of bandwidth, TxtVid achieves ultra-low bitrates of $\sim$$100$ bps by compressing videos to text. This can lead to multiple orders of magnitude compression advantage if we can achieve similar QoE as the conventional approaches in the low-bitrate regime.

There has been recent interest in the generative modeling community in reconstructing videos from lower bitrate alternatives such as text or low dimensional latent spaces \cite{li2018video, kim2020tivgan, weissenborn2019scaling, yang2018text2video}. While we see significant progress in using generative machine learning to model natural images from text \cite{ramesh2021zero, cho2020x, mansimov2015generating, yuan2019ckd, li2020exploring}, these approaches are currently unable to produce high-quality videos. To recreate webcam video data, 2D \cite{siarohin2019first, wang2018video, zakharov2020fast, prajwal2020lip, zhou2020makelttalk, eskimez2021speech, yu2021multimodal, choi2005automatic} or 3D graphics based methods \cite{suwajanakorn2017synthesizing, fried2019text, deng2020disentangled, cosatto2000photo} have been used successfully to generate realistic talking-head videos. Their success implies that machine learning methods have the potential to be used as decoders, that can reconstruct a webcam video with high QoE while requiring less data to be transmitted compared to standard codecs. Another recent line of work on achieving lower bitrates for talking-head videos has focused on using a source image to encode facial data and a driving video consisting of facial keypoints to encode the dynamic components of the video \cite{wang2020one, prabhakar2020reducing}. In this work, we take this approach one-step further by more aggressively compressing the talking-head videos into text, instead of a visual representation such as facial keypoints. 
We ask the question:\\\textit{``Can we compress AV content generated via webcams to just text and recover videos with similar QoE compared to standard codecs in a low bitrate regime?''}\\ and using recent advances in deep learning based generative models we answer it in the affirmative. The contributions of our work are as follows:

\begin{itemize}
    \item We propose a novel compression pipeline by compressing audio-video ``talking-head" videos to just text. The pipeline uses a state-of-the-art voice cloning model \cite{resemble} to convert text-to-speech (TTS), and a lip-syncing model \cite{prajwal2020lip} to convert audio to reconstructed video using a driving video at the decoder (Figure \ref{fig_bd}).
    \item We conducted a subjective study and our results demonstrate that at similar QoE in a low bitrate regime, the pipeline exhibits up to $100-1000\times$ compression advantage over standard audio-video codecs. 
    \item Information theoretically, our results can be viewed as establishing an empirical ``achievability'' result, showing that a rate of  $\sim$$100$ bps can yield reconstruction qualities, as assessed by humans, commensurate with what existing codecs would require orders of magnitude higher rates to achieve.   
    In particular, what we achieve is up to two orders of magnitude lower bitrate compared to results reported using facial keypoints \cite{wang2020one, prabhakar2020reducing}.
\end{itemize} 

Our pipeline can be used for storing the webcam AV content as a text file or for streaming this content on the fly. We envision the Txt2Vid framework can open up many novel application possibilities such as enabling audio-video communication during poor internet connectivity, or in remote terrains with limited bandwidth, including lunar or Martian space stations. It can be used for storing pedagogical content as text, and using the proposed decoder to learn from one's favorite instructor or disseminating information in multiple languages though text translation without the need to re-record video content. The code for our pipeline along with some examples and demonstrations is available on GitHub\footnote{https://github.com/tpulkit/txt2vid.git}, and the generated dataset for the subjective study is available to download on Google Drive\footnote{\url{https://tinyurl.com/33aadk6c}}.

\section{Compression Pipeline}
\label{sec_methods}

\begin{figure}
\centering
\includegraphics[width=\linewidth]{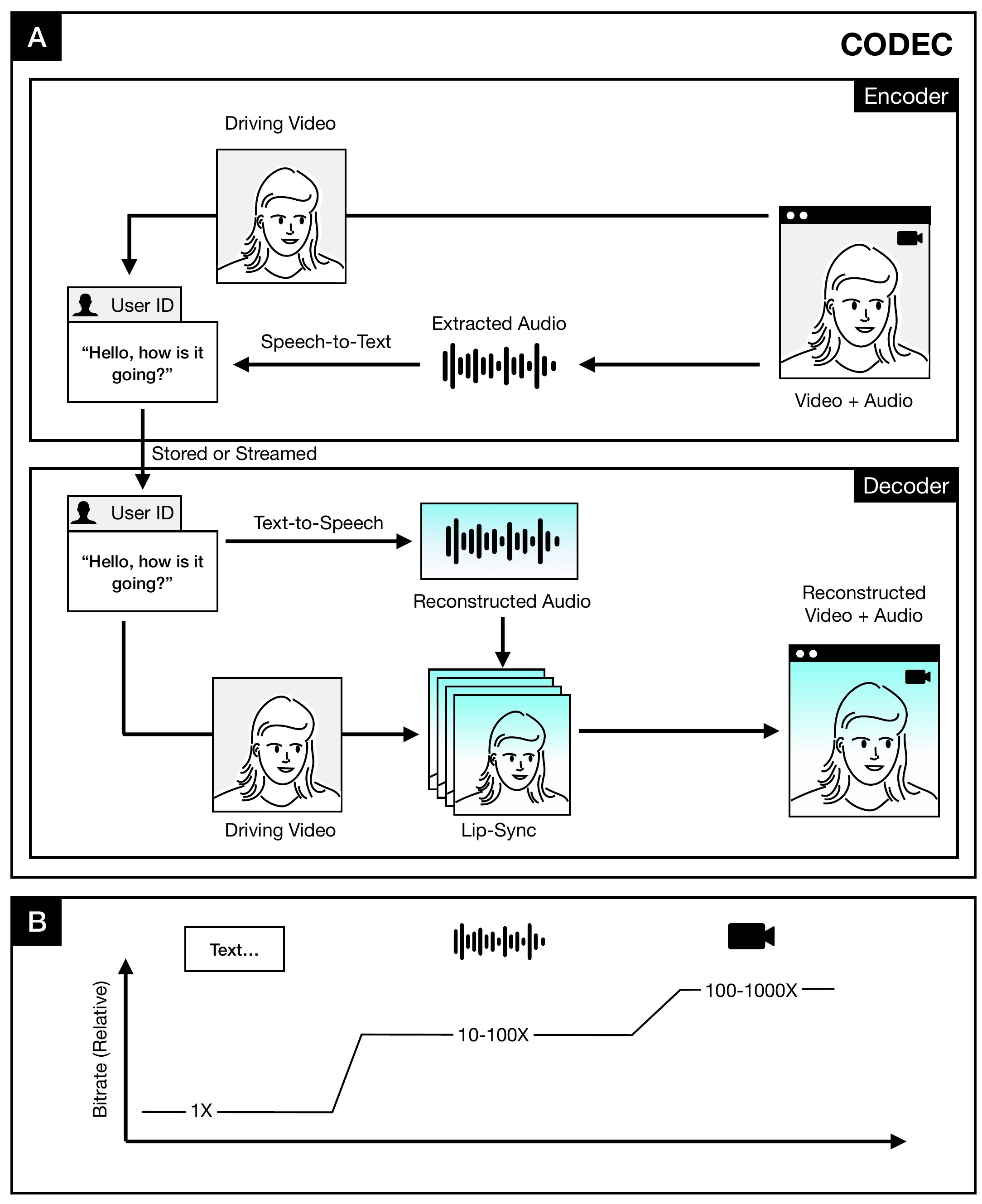}
\caption{Compression pipeline. (A) Block diagram of our codec. Text (or audio) is extracted from the original webcam video and transmitted. An additional driving video of the sender is transmitted once during the lifetime of communication between a sender and receiver pair, along with a ``User ID''. A generative text-to-speech model is used to recover audio from the text, followed by lip-syncing with the driving video to generate the final reconstructed video with audio output. (B) Expected compression advantage of transmitting text (or audio) over videos compressed using traditional methods.}
\label{fig_bd}
\end{figure}

The proposed pipeline is shown in Figure \ref{fig_bd} as a block diagram with focus specifically on video content involving a single person speaking in front of their camera as described in Section \ref{sec_intro}. The decoder takes the text transcript of such a video as input and outputs a reconstructed video with audio. This is done by first converting the text component to audio using TTS synthesis resulting in reconstructed audio, followed by speech-to-video (STV) synthesis. STV is done by lip-syncing the generated audio with a driving video (of the specific person in the original video content) available at the decoder. The driving video needs to be transmitted only once during the lifetime of communication between a particular sender and receiver pair. It is agnostic to the content of the current transmission, and can be extremely short ($\sim$$30$s) since playback can be looped. Thus, at the receiver, the driving video can be obtained prior to decoding the text message from a particular sender. Note that there can be multiple driving videos available at the decoder corresponding to different senders. Therefore during typical operation, a ``User ID'' to identify the appropriate driving video (and corresponding voice profile for reconstructed audio) needs to be transmitted. The driving video can be ignored when calculating transmission rate because it needs to be transmitted only once for a particular sender-receiver pair. 

The encoder takes as input the recorded webcam video and outputs the text transcript. Text can be extracted from the webcam video by using either automatic speech recognition (ASR)/speech-to-text (STT) tools or by manually transcribing the spoken content into a text file. This text can be further compressed using a standard compressor such as gzip \cite{gzip} or bzip2 \cite{bzip2}. The pipeline also allows audio to be transmitted, instead of text, which can help get better reconstruction fidelity at the cost of higher bandwidth usage. The pipeline covers the following modes of operation with increasing complexity: 
\begin{enumerate}
    \item Encoded text file is generated offline and the decoder acts as a video player. This enables storage of the content.
    \item Encoded text file is streamed, requiring real-time encoding and decoding but with some latency allowed. This enables applications like web-streaming where latency of the order of $\sim$$5$s is acceptable \cite{wowza2020}.
    \item Encoded text file is live-streamed. This requires a tight bound on latency along with real-time encoding and decoding. This mode allows interactive real-time communication such as video calls between participants.
\end{enumerate}

For performance evaluation (Section \ref{sec_pe}), we modified a pre-trained lip syncing model, Wav2Lip \cite{prajwal2020lip} and used the Resemble API \cite{resemble} for TTS. Our main focus was to demonstrate the capability of the proposed pipeline to drastically reduce the data transmission rate while maintaining QoE. Therefore, for performance evaluation we limited ourselves to the decoding pipeline which generates a video file from a text transcript. However, in practical systems, one may also be interested in streaming applications. To that end, we built an additional prototype to demonstrate how our approach can be used as part of a video streaming system. Details about how the various software tools were utilized and modified for this work are provided in the Appendix \ref{sec_software_tools}, along with demonstrations of the streaming prototype on our Github page.

As shown in Figure \ref{fig_bd}B, the pipeline has the potential to achieve extremely low bitrates, up to $100$-$1000\times$ smaller than videos compressed using standard codecs. In fact, text can be communicated at $\sim$$100$bps, whereas current codecs cannot achieve such high compression even at extreme settings. The current implementation can lead to loss or alteration of details like facial expressions, tone of voice, and prosody of speech. The QoE can be improved substantially in future versions by incorporating models which also account for these factors, at a cost of relatively small bandwidth increment required for transmitting the additional metadata.

\section{Performance Evaluation}
\label{sec_pe}

In this section, we evaluate the compression gains and reconstruction quality achieved by the proposed pipeline. The evaluation was performed through a subjective study with 242 participants. The study involved comparing videos reconstructed from a text transcript against the original videos encoded using different standard codec parameters. The evaluation focused on understanding the bitrate and QoE achievable by the pipeline, and hence it did not include the streaming mode or the STT-based encoder. We first discuss the dataset used for evaluation and present details on the subjective study for comparing standard codecs against the proposed approach, followed by the results.  

\subsection{Dataset}
\label{sec_pe_ds}

\begin{figure}
\centering
\includegraphics[width=\linewidth]{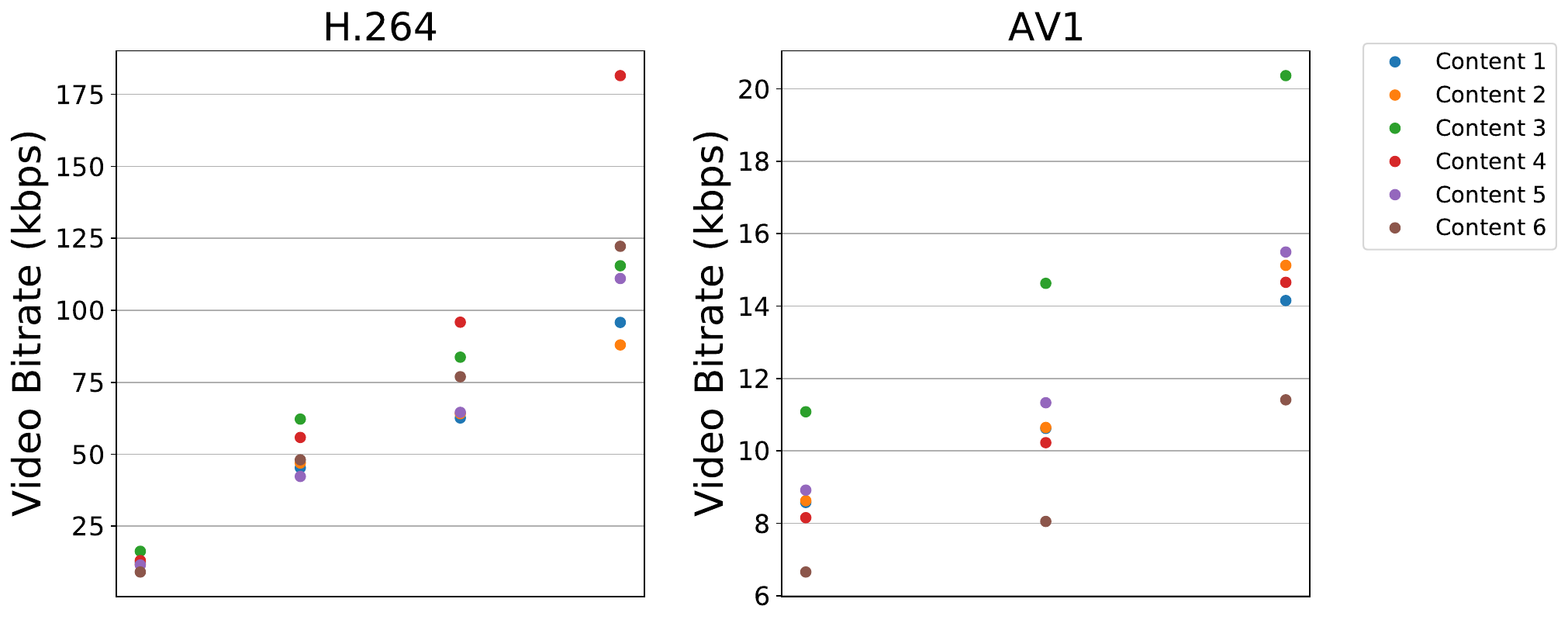}
\caption{Benchmark Dataset Video Bitrates. Left and right panels show video bitrates for the six contents encoded using H.264 and AV1 as the video codecs. For each content, we encode at 4 different H.264 and 3 different AV1 configurations. Audio for each of these configurations was encoded at two different bitrates using AAC. This results in total of $4\times2 + 3\times2 = 14$ different benchmark AV per content. Details of the codec parameters used to generate the benchmark dataset are available in Appendix \ref{app_dataset_gen}.}
\label{fig_ds_props}
\end{figure}

A dataset was created for the study by recording webcam talking-head videos. Six $\sim$$30$s videos were recorded by six different people (different ethnicities; four male, two female) under diverse natural indoor ambient lighting and speaking conditions to serve as the original AV content. Each video consisted of the speaker talking about a different technical topic. These videos were used to create several AVs using standard codecs (benchmark set) and our approach (Txt2Vid set). The generated dataset is available on Google Drive (link in Section \ref{sec_intro}), and exemplary encodes are shown in Figure \ref{fig_examplary_encodes} (Appendix \ref{exemplary_encodes}).

The benchmark set was generated through ffmpeg using the following steps: 1) convert original video to 720p resolution, 25 fps and yuv420p, 2) encode the video using H.264 or AV1 codec, at a particular encoding parameter, viz.\ CRF (constant rate factor) and downsampling ratio, 3) convert original audio to sampling rate of 16kHz, 4) encode audio at a constrained bitrate (BR) using AAC codec, 5) merge encoded audio and video. We observed that since we are working with extremely low bitrates, a better quality at similar bitrates is achievable by first downsampling the video (audio), followed by using a better quality parameter for compressing. This compressed and downsampled video (audio) can then be transmitted and recovered at desired parameters by upsampling at the decoder. Therefore, for each reported bitrate in the dataset, we tried achieving similar bitrate by varying CRF (BR) along with downsampling by $1\times$, $2\times$ and $4\times$  ($1\times$, $2\times$) for video (audio). We chose the downsampling setting which provided the best quality for the subjective study. We chose 4 different encoding properties for H.264 and 3 different ones for AV1. Two different encoding properties of AAC resulted in a total of $14$ different encodings per video content, and a total of $84$ benchmark videos. Obtained audio-video bitrates are shown in Figure \ref{fig_ds_props} and details of the codec parameters used are given in Appendix \ref{app_dataset_gen}. The choice of the encoding parameters was also informed by a small pilot study we conducted before the main study to gauge the appropriate range of bitrates to work with. Diversity in video bitrates at similar CRF across contents as seen in Figure \ref{fig_ds_props} highlights the diversity present in the dataset.

The Txt2Vid set was generated by utilizing a bzip2 compressed text transcript file from the original content \cite{bzip2}. This text transcript was first converted to audio using the voice clone from the Resemble API for each individual in the dataset, followed by using Wav2Lip for lip-syncing with an independent driving video as described in Section \ref{sec_methods}. The driving video was speaker specific but agnostic to the spoken content. The driving video was encoded using H.264 at CRF of $20$, 720p resolution, 25 fps and yuv420p. The Txt2Vid set also contained videos generated using Wav2Lip by directly passing audio through the pipeline instead of the reconstructed audio from Resemble. The audio used was encoded using AAC at $\sim$$10$kbps and at a sampling rate of 16kHz. Overall, the Txt2Vid set contains $2$ videos per video content resulting in total of $12$ Txt2Vid videos.

\begin{table}
\caption{Bitrates of Txt2Vid Videos generated from text}
\begin{tabular}{c|c|c|c|c|c|c}
Content               & 1 & 2     & 3     & 4     & 5     & 6 \\ \hline
Txt2Vid Bitrate bps & 79.20 & 81.33 & 92.27 & 83.47 & 82.67 & 89.6                  
\end{tabular}
\label{table_txt2vid_brs}
\end{table}

\begin{figure*}[htbp]
\centering
\subfloat[Using H.264 video codec + AAC audio codec.]{\includegraphics[width=0.49\textwidth]{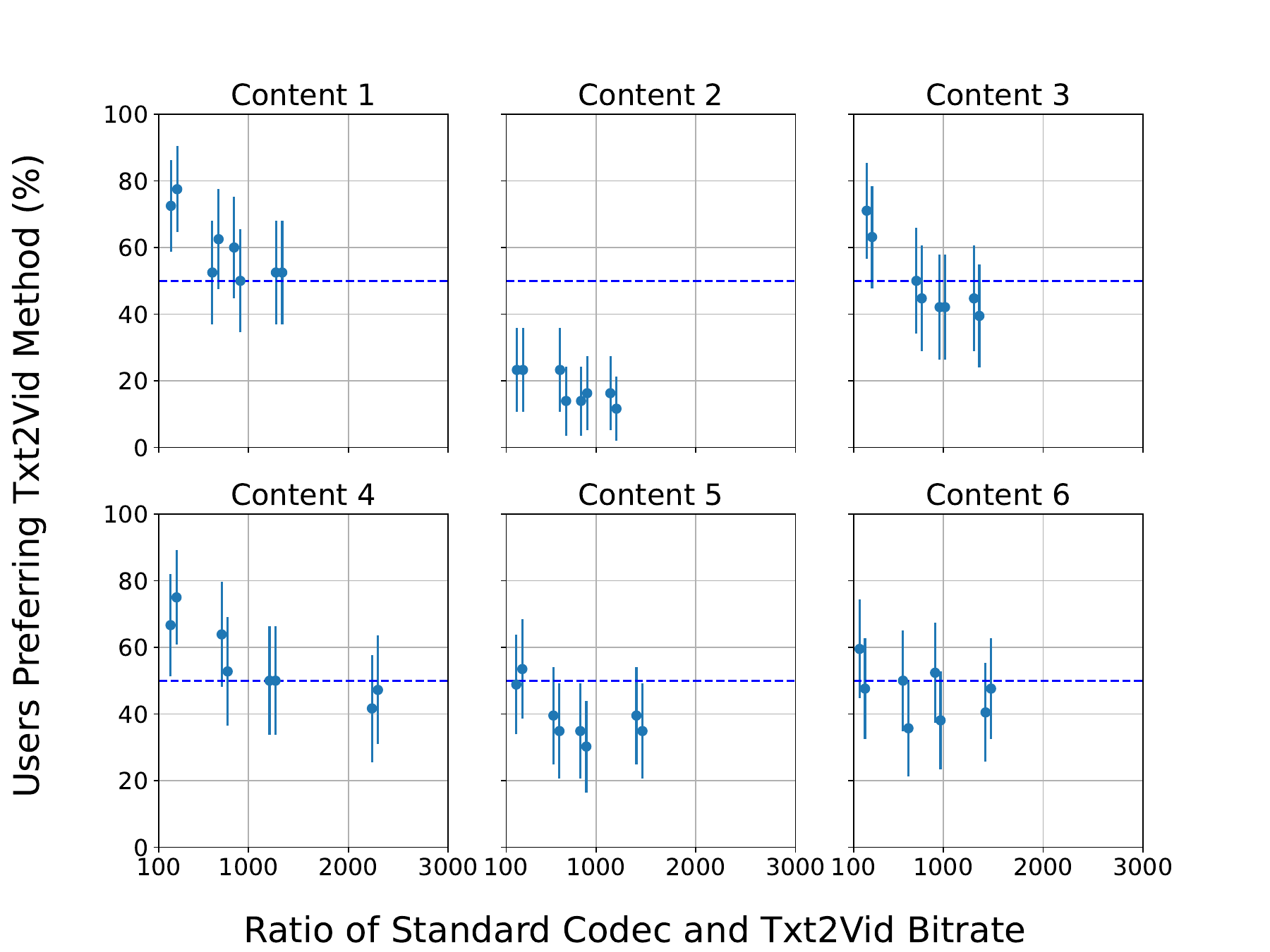}\label{fig_results_avc}}
\subfloat[Using AV1 video codec + AAC audio codec.]{\includegraphics[width=0.49\textwidth]{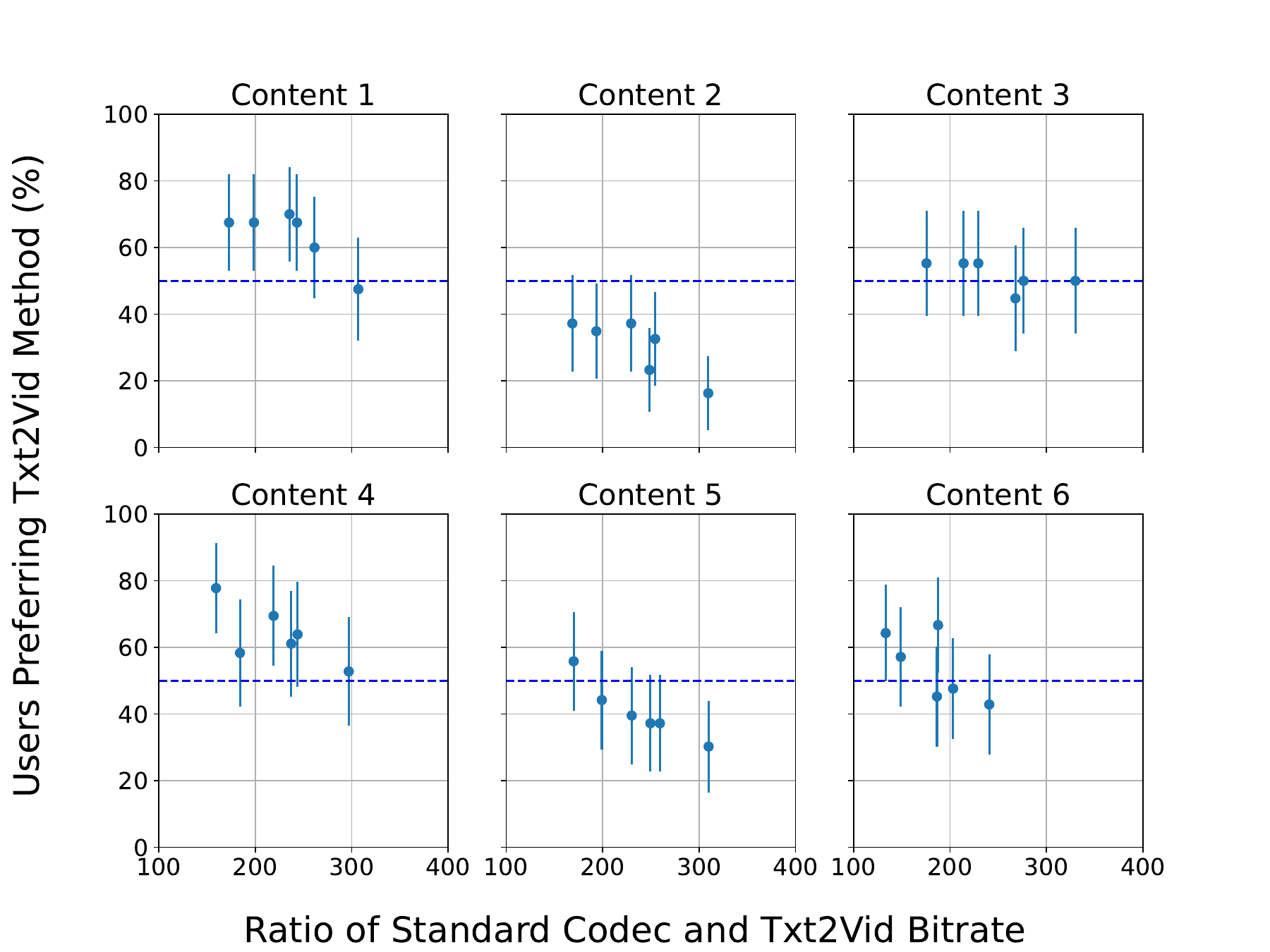}\label{fig_results_av1}}
\caption{Subjective Study Results: comparing Txt2Vid videos generated from text to standard codecs. Each plot shows the percentage of users who prefer the Txt2Vid video over the standard codec vs.\ the ratio of the bitrate between the standard codec and Txt2Vid. We compare Txt2Vid against both (a) the widely used H.264 video codec, and (b) state-of-the-art AV1 video codec. Both (a) and (b) use AAC as the audio codec as it is most widely used and considered state-of-the-art. The dashed horizontal line represents the 50\% preference level where the two methods are equally preferred by the study participants. The error bars show 95\% confidence interval.}
\label{fig_resemble_avc_av1}
\end{figure*}

\begin{figure*}
\centering
\subfloat[Using H.264 video codec + AAC audio codec.]{\includegraphics[width=0.49\textwidth]{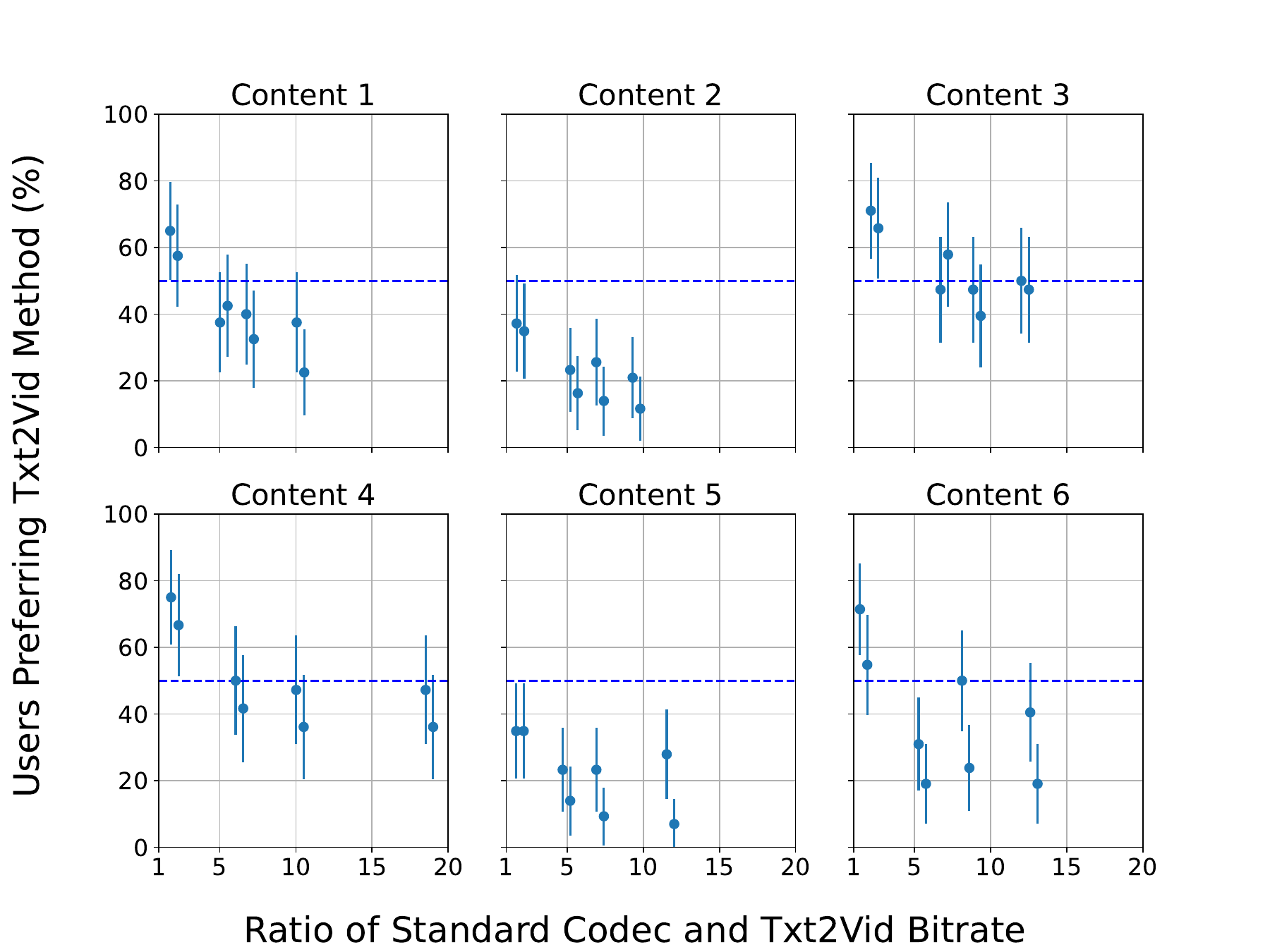}\label{fig_results_audio_avc}}
\subfloat[Using AV1 video codec + AAC audio codec.]{\includegraphics[width=0.49\textwidth]{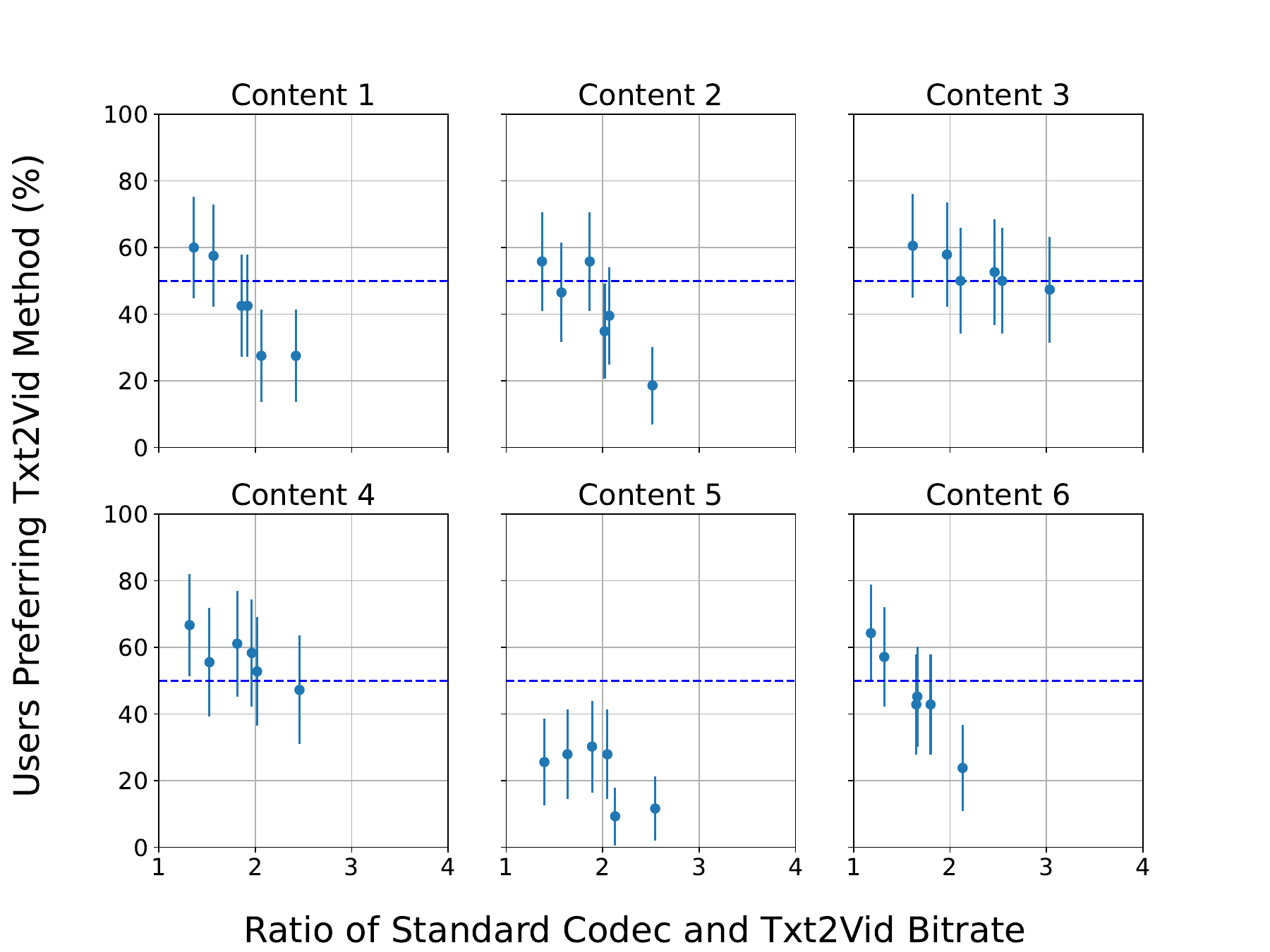}\label{fig_results_audio_av1}}
\caption{Ablation Study: subjective study results comparing Txt2Vid videos generated from original audio to standard codecs. Each plot shows the percentage of users preferring the Txt2Vid video over the standard codec vs.\ the ratio of the bitrate between the standard codec and Txt2Vid. The dashed horizontal line represents the 50\% preference levels where the two methods are equally preferred by the participants. The error bars show 95\% confidence interval.}
\label{fig_orig_avc_av1}
\end{figure*}

\begin{figure}
\centering
\includegraphics[width=\linewidth]{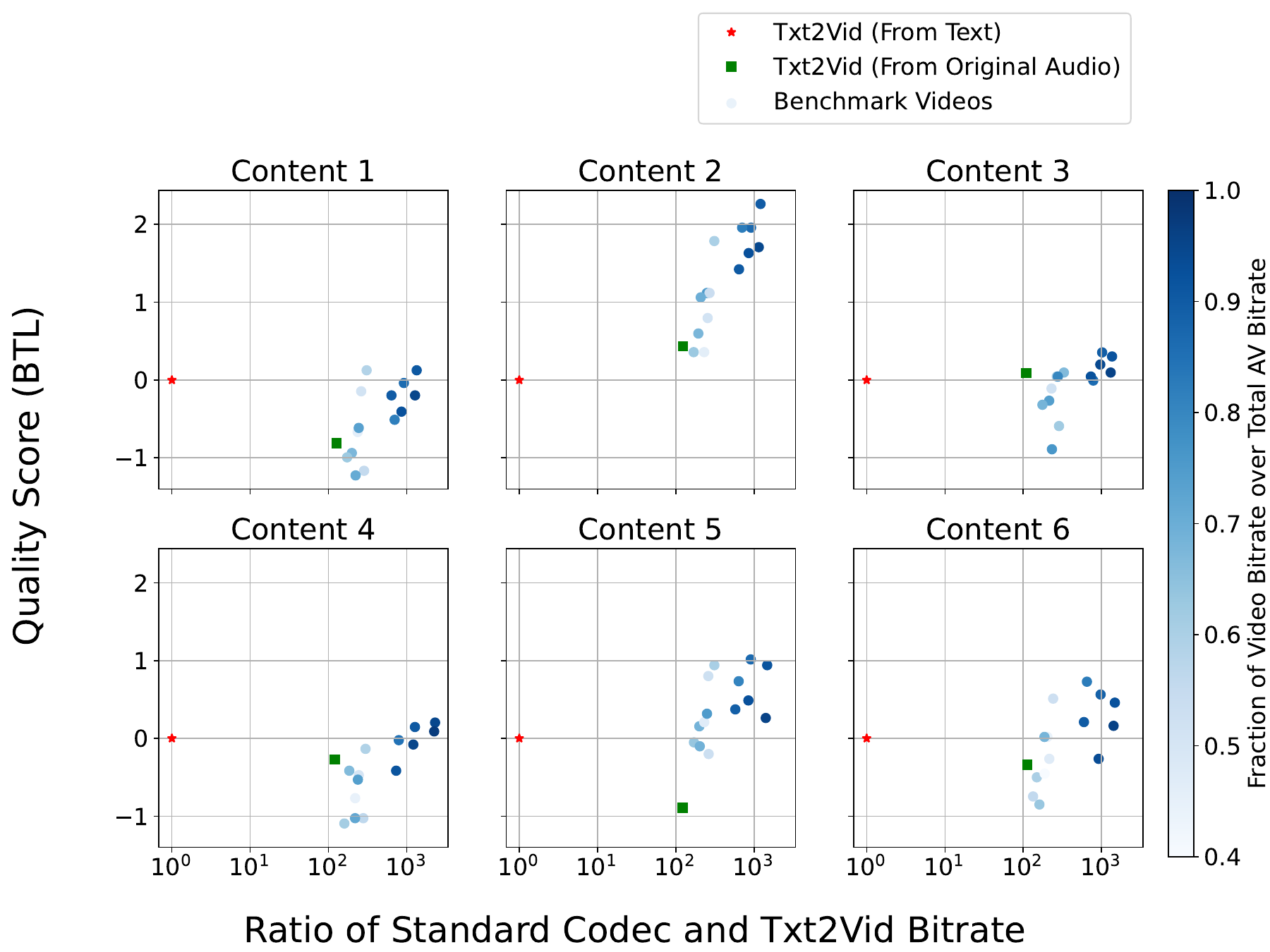}
\caption{Quality Scores using BTL model. Each plot shows the inferred quality scores on an interval-scale inferred by modeling complete paired comparison data using probabilistic Bradley-Terry-Luce model \cite{chen2009crowdsourceable}, against the ratio of the bitrate between the standard codec and Txt2Vid. Txt2Vid videos (red star) serve as reference with quality score of 0, and a positive score implies a higher inferred quality for the corresponding video. This demonstrates potential of the proposed Txt2Vid approach to achieve two-to-three order of compression over standard codecs. The color bar for the benchmark videos shows the fraction of total AV bitrate spent on the video component.}
\label{fig_bt_results}
\end{figure}

\begin{figure*}
\centering
\includegraphics[width=0.8\textwidth]{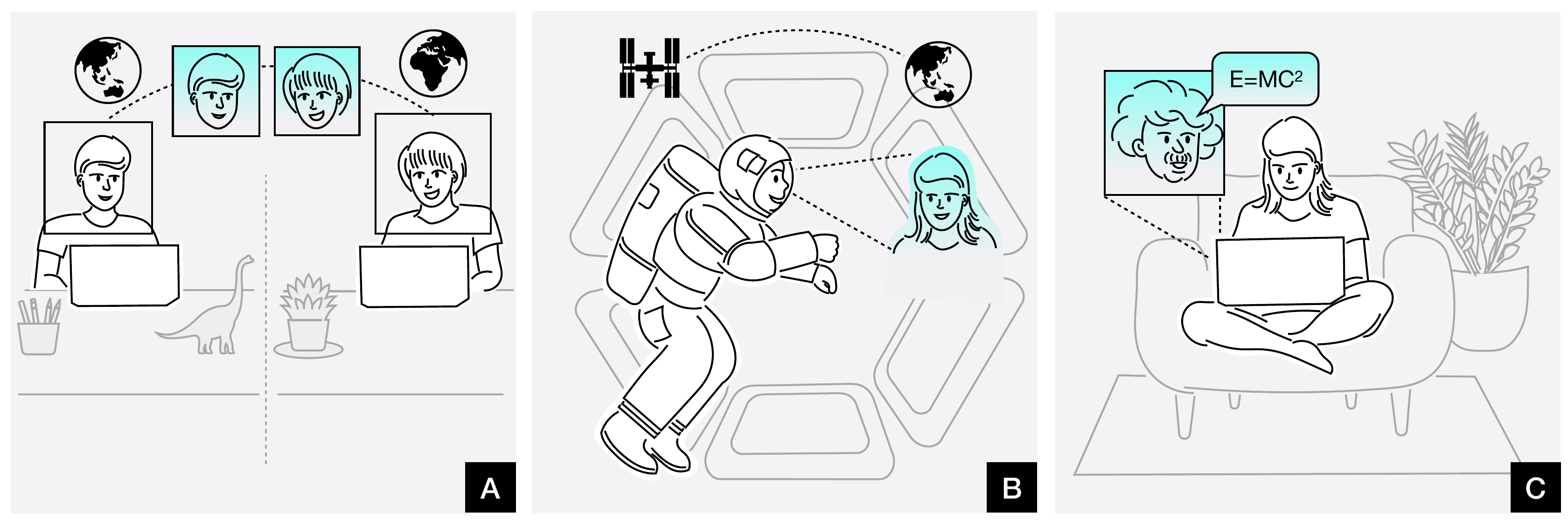}
\caption{Applications. Txt2Vid framework opens up novel applications such as: (A) enabling AV communication during poor internet connectivity, (B) enabling AV communication in remote terrains with limited bandwidth, or (C) storing pedagogical content as text and using it to learn from your favorite instructor.}
\label{fig_applications}
\end{figure*}

\subsection{Subjective Study}
\label{sec_pe_ss}
The subjective study\footnote{IRB Number: E-3152} was generated using Qualtrics Platform and conducted using Amazon MTurk. MTurk workers were required to have ``masters'' qualification and lifetime approval rate greater than 98\% to be qualified for our study. For each video content (six recordings in our dataset), we did a pairwise comparison test of all videos in the benchmark set with each video in the Txt2Vid set resulting in $14\times2=28$ comparisons per content. Thus, a total of $28\times6=168$ pairwise video comparisons were subjectively studied. Since watching many video pairs of $30$s each can be time consuming and tiring for a study participant, we decided to show only $28$ comparisons to each viewer belonging to one video content (and two additional comparisons for sanity check with obvious audio or video degradation respectively). Each pair was compared by $\sim$$40$ viewers leading to a total of $252$ participants. Each video in the pair was shown at a resolution of 480p, and the viewers were asked to choose which video in each pair they preferred. We asked a general preference question instead of a specific quality question such as ``which video has a higher quality'' to account for varying personal preferences amongst viewers across video and audio qualities. We manually verified responses from the participants for ensuring quality of the responses and removed $10$ participants from the study who either completed the study too fast ($<10$min) or failed both the sanity checks. Our results in next section are reported over the remaining $242$ participants. Screenshots from the subjective study test are also available on the Google Drive link provided above.

\subsection{Results and Analysis}

\label{sec_pe_results}
\subsubsection{Achievability of talking-head video communication at bitrates as low as $\sim$$100$ bps}
In comparison to the benchmark dataset with bitrates ranging from $\sim$10-100 kbps, the proposed Txt2Vid method requires only $\sim$85 bps on average across all contents evaluated in the study (Table \ref{table_txt2vid_brs}). This is relatively close to $\sim$39 bps information rate empirically estimated for spoken communication in many languages \cite{coupe2019different} and serves as an empirical lower-bound on the bitrate required for meaningful talking-head video communication. The difference between Txt2Vid and spoken language encoding rates can be partly explained by the shorter text segments in the study which are harder to compress using bzip2. \\

\subsubsection{Txt2Vid achieves two-to-three orders of magnitude compression at similar QoE compared to standard codecs}

Figure \ref{fig_resemble_avc_av1} shows the results for the subjective study comparing the Txt2Vid generated reconstructions (from text) with the standard codecs (AVC and AV1 video codecs shown separately) for all six contents. Recall that we used two audio bitrates (AAC codec) for each video codec setting, both of which are shown in the plots. Note that only the set of videos generated from text using Resemble API are shown here, while the results for the set of videos generated by directly using the original audio are provided in Section \ref{sec_ablation_results} (Ablation Study). The plots show the percentage of users preferring the Txt2Vid method over a given standard codec setting against the ratio of bitrates for the standard codec and Txt2Vid. As this ratio increases, the reconstruction quality for the standard codec improves and we expect fewer people to prefer Txt2Vid. We observe this monotonicity in the plots, along with some outliers to the trend. These outliers can be explained by statistical noise and the fact that we can get different overall qualities at the same bitrate depending on how the total bitrate is allocated to video and audio. The error bars show the 95\% Confidence Interval using standard normal distribution assuming each choice can be modeled using a binomial distribution. Focusing on the 50\% preference level (dashed line in plot), we observe that for most contents Txt2Vid achieves similar QoE with up to $1000\times$ smaller bitrates than the widely used H.264+AAC codec (Figure \ref{fig_resemble_avc_av1}a). Even when using the state-of-the-art AV1 video codec, we still see close to $200\times$ lower bitrates using Txt2Vid at similar user preference (Figure \ref{fig_resemble_avc_av1}b). These results illustrate the promise of our pipeline in reducing videos to a minimal text-based representation followed by generative reconstruction. Figure \ref{fig_examplary_encodes} in Appendix \ref{exemplary_encodes} shows frames belonging to exemplary Txt2Vid and baseline encodes found at similar-QoE using our subjective study for a few different contents in the dataset.

\subsubsection{Ablation Study}
\label{sec_ablation_results}
To discern the advantages coming from the whole pipeline against just the lip-syncing, we conducted an ablation study. Instead of using TTS audio, we passed the original audio into the lip-syncing decoder. 
Figure \ref{fig_orig_avc_av1} shows the results for the subjective study comparing the Txt2Vid videos generated from original audio (at 10kbps) against the standard codecs. The overall trend remains similar to Figure \ref{fig_resemble_avc_av1} but the reduction in bitrate for Txt2Vid over standard codecs at the 50\% preference level is much smaller. We achieve $\sim$$5\times$ reduction over AVC and $<$2x reduction over AV1. The audio bitrates are comparable to video bitrates in the low bitrate regime, and hence we do not save as much by just transmitting the audio. An attempt to reduce the audio bitrate further in our study resulted in discernible artifacts in the encoded audio, resulting in much worse QoE.  

\subsubsection{Inference of Quality Scores from Subjective Study}
\label{sec_btl_model}
To further elucidate our results, we modeled the obtained QoE results with the widely-used Bradley-Terry-Luce (BTL) model \cite{bradley1952rank, luce2012individual, chen2009crowdsourceable}. BTL converts the obtained paired comparison results to an intrinsic quality score for each video on an interval-scale. BTL uses a logistic regression model: assume $Q_i$ is the intrinsic quality score of video $i$ and $P_{i,j}$ is the probability assigned by model that a user would prefer video $i$ over video $j$, then 
\begin{equation}
    P_{i,j} = \frac{e^{Q_i - Q_j}}{1 + e^{Q_i - Q_j}}
\end{equation}

Let $frac(i>j)$ be the fraction of users preferring video $i$ over video $j$ in the observed data, then the quality scores $Q_i$ for each video can be inferred by maximizing the likelihood of $frac(i>j)$ given model probabilites $P_{i,j}$ through an optimization procedure such as the Newton-Raphson method.

Figure \ref{fig_bt_results} shows the obtained Quality Scores using BTL model against the ratio of bitrates for the standard codec and Txt2Vid, for each content separately. These quality scores were extracted by doing MLE over all paired comparisons across 16 videos used in our subjective study (per content): [($4$ AVC + $3$ AV1) $\times$ ($2$ AAC)] + [Txt2Vid (from Text) + Txt2Vid (from Original Audio)]. Since the logistic model stays the same under addition of a constant over the quality scores, we report all quality scores with respect to Txt2Vid generated video (shown as a red star in the figure). Thus, a positive quality score implies a higher inferred quality of the video compared to TxtVid, and vice-versa. As the ratio of standard codec to Txt2Vid bitrate increases, we expect the reconstruction quality and obtained quality scores to be higher and we see that in the plots. Figure \ref{fig_bt_results} shows that our Txt2Vid approach provides two-to-three orders of magnitude compression at similar quality scores. 

Surprisingly, we found that for most of the contents, Txt2Vid videos reconstructed from text using generated audio had higher quality scores than Txt2Vid videos reconstructed using original audio. This shows that at extremely low bitrates, audio quality likely starts dominating the user's experience. However, even at a few kbps, compressed audio quality for spoken content is relatively poor. TTS alleviates this issue by regenerating high quality audio samples at the decoder, without requiring transmission of original audio. Thus, the orders of magnitude advantage we report in this work is coming from generative modeling of both video and audio together. 


\subsubsection{Analysis across content}
We note that there is some variation between the results for the different contents which is expected due to the diversity in the background and lighting conditions, e.g. for some contents we observed that the lip synced region was visible as a rectangular artifact making the video appear unnatural, as well as voice cloning quality. In particular, the preference for Txt2Vid is lower for content 2 (and content 5 for the case of Txt2Vid videos generated from original audio). Note that a non-negligible fraction of participants still prefer Txt2Vid over the standard codecs which require orders of magnitude higher bitrate. To better understand the performance variation, we analyzed the comments from participants, and found that some of them listed audio quality as the primary determinant for their preferences. For most contents, the audio produced by Resemble has much better quality as compared to the benchmark set (which had audio with relatively low bitrates). However, in certain cases, the Resemble audio sounded unrealistic, robotic and/or lacked clarity. This was also observed in a comparison between Txt2Vid with Resemble audio and Txt2Vid with original audio (at 10 kpbs bitrate), where only 30\% participants preferred the Resemble audio for content 2 (in contrast to more than 50\% preference for most other contents). We believe that this is an artifact of the quality of training data and the Resemble training process, and we expect things to improve as voice cloning technology progresses. For the other contents, this highlights a major advantage of the proposed pipeline - the ability to obtain high quality audio at extremely low bitrates using voice cloning and TTS.

\section{Discussion}
\label{sec_discussion}

\subsection{Applications}
We believe that our setup can enable several applications with positive societal impact as shown in Figure \ref{fig_applications}. Due to the extremely low bandwidth requirements, this can allow people in areas with poor internet availability, high costs and limited access, including those living in remote areas and in the developing world, to be better connected and get a good audio-video experience. This pipeline is particularly suited for transmission of pedagogical content, which is topical given the growth in remote learning and online instructional videos. Since the pipeline only transmits a text transcript, it also opens up the possibility of using different voices and faces to help students feel more engaged with the content. One can imagine Albert Einstein teaching relativity or a child's favorite movie character teaching them math. Given advances in machine translation, this system can be easily modified to display videos in multiple languages without the need to create and store multiple original versions. 

Even for a normal video calling application, this work opens up many possibilities. A user can speak as usual and the communication can occur via the transcribed text or the original audio. The reconstructed video seen by the other users can potentially use any face or voice, no longer subject to the various implicit biases. A user caught up in some other task can simply type in what they wish to say, without affecting the audio-video experience for the other users. We plan to further investigate these applications.

\subsection{Limitations and Future Work}
\subsubsection{Computational Complexity}
The proposed pipeline is currently a prototype to demonstrate the advantages of a generative approach. More work is needed to enable widespread use in daily life especially in the context of streaming. The high computational complexity of lip syncing and the requirement for GPUs is a bottleneck for using this system on low-end devices which are more likely to be found in regions with poor connectivity. In addition, the current reliance on cloud-based APIs for TTS and STT is not well-aligned with the broader aim of reducing bandwidth usage. But, with existing high investment in hardware to accelerate deep learning models, reduction in GPU costs, significant research in edge-computing, and more efficient open-source models, we envision that these limitations will be considerably alleviated in the near future. As part of future work, we hope to build a standalone application on desktop and mobile devices to make this system widely accessible.

\subsubsection{Latency}
The latency of the streaming pipeline needs to be reduced to enable real-time interactive applications such as video calling. As described in Section \ref{sec_methods_streaming}, the streaming system currently is a proof-of-concept rather than a production-level system. The current end-to-end latency is close to 4 seconds, largely due to the buffering at ffmpeg. We believe that this can be reduced by using a real-time protocol, and we plan to explore this in future.

\subsubsection{Quality-of-Experience}
In addition to the above technical limitations, there is scope for improvement in the encoding and reconstruction quality. We found that the STT occasionally incorrectly transcribes certain words, especially technical words, and requires manual proofreading. Furthermore, the system by design is restricted to communicating the transcript, and thus can alter verbal content (e.g., tone of voice, prosody of speech) or miss non-verbal cues (e.g., head nods, eyebrow raises), non-speech sounds (e.g., laughter), and facial expressions. We note that this is not a fundamental limitation and one can envision a system that transmits additional metadata along with the transcript to capture these aspects, coupled with a decoder capable of incorporating them into the reconstruction. This would involve training newer models that integrate these non-verbal communication features and is left as future work.

\subsubsection{Ethical and Privacy Considerations}
Finally, we note that a pipeline like ours raises some privacy concerns and has a potential for misuse typically associated with Deepfakes \cite{vaccari2020deepfakes}. For example, transmitting video as text allows for any face or voice to be used at the decoder potentially allowing misportrayal of an individual's identity. A mechanism to limit the usage of generative models to the duration of the call could potentially abate the misuse. One way to do so would be to integrate a security mechanism (such as an encrypted key) at the receiver which only allows generating content using the sender's identity when explicitly allowed by the sender, or limited to the duration of the call. New ways to distinguish actual audio/video from the generated audio/video, both at a human and computational level are required so that users at all times are aware they are interacting with generated video. In summary, widespread usage of this technology requires cooperation between governments, industry and academia to develop safeguarding mechanisms and address these challenges at legal, technical and societal levels \cite{leibowicz2021deepfake}.

\section{Conclusion}
In this work, we presented a novel video compression pipeline Txt2Vid, for extreme compression of talking-head videos as seen in video conferencing and webinars. Txt2Vid minimizes the data transmission rate by reducing the video to a text transcript, followed by a realistic reconstruction of the video using recent advances in deep learning based voice cloning and lip syncing models. We implemented a prototype using Resemble voice cloning and Wav2Lip lip-syncing frameworks. We also demonstrated a proof-of-concept streaming pipeline that can potentially enable real-time applications in the future. We evaluated our pipeline using a subjective study on Amazon Mturk to compare user preferences between Txt2Vid generated videos and videos compressed with standard codecs at varying levels of compression. In the study, performed on multiple video contents, our proposed pipeline achieved two to three orders of magnitude lower bitrates than state-of-the-art audio-video codecs at comparable Quality-of-Experience. The proposed framework can enable several applications with great potential for social good, expanding the reach of video communication technology. While we used specific tools in our pipeline to demonstrate its capabilities, we envision significant progress in the components used over the coming years leading to even better reconstruction quality. 

\section*{Acknowledgment}
The authors would like to thank all the participants of our subjective user study. The Stanford authors have been partially supported by Meta (formerly Facebook).

\ifCLASSOPTIONcaptionsoff
  \newpage
\fi


\bibliographystyle{IEEEtran}
\bibliography{citations.bib}

\appendices

\section{Software Tools} \label{sec_software_tools}

\subsection{Lip-syncing Model} \label{sec_lip_sync_model}
For STV synthesis, we focus on models for generating realistic talking-head videos from audio. These models are built on the observation that realistic talking-heads can be generated from audio by mostly capturing the mouth movement in the video. These models can be categorized into, (a) audio-driven 3D facial animation \cite{wang20213d, taylor2017deep, cudeiro2019capture, karras2017audio, zhou2018visemenet, fanelli20103}, or (b) lip-synced talking-head videos \cite{prajwal2020lip, chen2019sound, fried2019text, jamaludin2019you, kr2019towards, thies2020neural,zhou2020makelttalk, verma2004animating}. 3D models generate the whole frame by first generating an audio-driven 3D model of an individual followed by a 2D projection into a frame. Lip-syncing talking-head videos simplify the problem further by not generating the whole frame from just the audio. They utilize supplementary information from a driving frame or video. This driving frame (or video) is used as a prior for the model generating the synced lip-movements with the audio. This generation method reduces the computational complexity by only generating the region around the lip, and provides the best output quality as a lot of visual information is already contained in the driving frame (or video). We work with a pre-trained lip-syncing model -- Wav2Lip \cite{prajwal2020lip}.

Wav2Lip takes as input the driving video (or frame) and an audio clip, and outputs a lip-synced video or frame with the audio. In our case, the driving video is a short silent clip ($\sim$$30$s) of the person facing the webcam with natural head/eye movements. Wav2Lip relies on a generative adversarial network (GAN) trained with the help of a pre-trained discriminator that can detect lip-syncing errors. During inference, Wav2Lip first identifies the face and lip region in the driving video frames, and then uses the trained generator model to reconstruct this region. It has been shown to work well with dynamic and unconstrained talking-head videos. However the existing implementation requires the complete audio for generating a lip-synced video which does not work for our pipeline as the decompressor needs to decode the stored or streamed text files on the fly. We modified the code from Wav2Lip GitHub repository\footnote{\url{https://github.com/Rudrabha/Wav2Lip}} to enable the generation of lip-synced videos in a streaming manner. This was done by chunking the audio stream used to generate videos and ensuring a batch size of one frame during inference for frame-by-frame generation. Some caveats of using the existing Wav2Lip pre-trained model include, (a) the audio chunk size has to be greater than $200$ ms because of model architecture leading to a fixed minimum latency, and (b) the model works well only with single person driving videos with frame resolution of $720$p and the person directly facing the camera. These drawbacks are a consequence of using the Wav2Lip model which is trained on a specific dataset. They can be overcome by training a new lip-sync model from scratch on different datasets depending on the application use cases.

\subsection{Text-to-Speech} \label{TTS_model}
As compared to STV, TTS is a relatively mature technology with many commercial vendors providing an API such as Google \cite{googleTTS}, Microsoft \cite{microsoftTTS}, Resemble \cite{resemble}, and Descript \cite{descriptTTS}. TTS technologies use machine learning based generative modeling to create realistic audio samples from text. They allow for the generation of voice samples in an individual's voice (voice cloning) to more closely represent the original person. Thus, using currently available TTS technology provides a timely and unique opportunity for the proposed compression pipeline. In this work, we use Resemble API \cite{resemble} to generate natural voice clones of individuals and use them during inference for decoding. Training a voice clone on Resemble is fast and easy. It requires recording a minimum of $50$ predefined voice samples on their website. Our current prototype relies on a web-based API for TTS leading to additional communication overhead for sending and receiving the text script and the generated audio, respectively. But this is a constraint due to the presently available TTS systems, and we do not include this as part of the transmission rate since an actual production-level system could perform TTS locally \cite{baevski2020wav2vec}.

\subsection{Streaming Prototype}
\label{sec_methods_streaming}
Our main focus in this work was to demonstrate the capability of the proposed pipeline to drastically reduce the transmission rate while maintaining good QoE. Therefore, for performance evaluation (Section \ref{sec_pe}) we limited ourselves to the decoding pipeline which generates a video file from a text transcript. However, in practical systems, we may also be interested in streaming applications. These applications typically have a range of latency requirements depending on whether real-time interaction is desired. In addition, real-time video playback requires real-time decoding to prevent degrading the user's experience. Therefore, we built an additional prototype to demonstrate how our approach can be used as part of a streaming pipeline. 

At the encoder, we used Google STT streaming API \cite{googleSTT} to convert the speech audio to text in real-time. The API returns finalized text for the preceding speech segment after it detects a pause in spoken content. This text is then transmitted to the receiver using websockets. One limitation of the current implementation is that the latency varies depending on the length of the sentence ($\sim$1-10s), which can sometimes lead to unexpected silence on the receiver's end. The variability in latency can be managed with a buffer on the receiver though at the cost of increased latency.

At the decoder end, we receive text from the websocket or read it from a file, and send it to Resemble using their API (one paragraph at a time). A separate callback server receives the generated audio and sends it to the processing script through a pipe. The voice cloned audio is split into 200ms chunks (see Appendix \ref{sec_lip_sync_model}) and passed to the Wav2Lip generator. Finally, the generated frames and the audio are combined using ffmpeg and sent to an HTTP port for video playback using ffplay or VLC. The receiver can also accept audio directly from the websocket or from an audio file, in which case the TTS part is no longer needed and is skipped. We used multithreading and pipes/queues in Python to build this system\footnote{Demonstration videos along with detailed instructions are available on GitHub.}.

\section{Dataset Generation.}
\label{app_dataset_gen}
Table \ref{table_codec_params} enlists all the parameters used for the generation of 14 different encodes, referred to as benchmark set in Section \ref{sec_pe_ds}. Same parameters were used for all video contents. Codec-V (Codec-A), CRF, DS-V (DS-A), BR and Avg. Bitrate refer to the video (audio) codec, Constrained Rate Factor for video encoding, video (audio) downsampling, constrained bitrate for audio encoding, and average combined audio and video bitrate across all contents respectively. A higher CRF and higher video downsampling implies a lower video quality at lower bitrate. A lower BR and higher audio downsampling implies a lower audio quality at lower bitrate. The choice of parameters was driven by an attempt to get the best subjective quality at a given total bitrate. 

\begin{table}
\caption{Codec Parameters}
\begin{center}
\begin{tabular}{|c|c|c|c|c|c|c|}
\hline
 \multirow{3}{*}{Codec-V} & \multirow{3}{*}{CRF} & \multirow{3}{*}{DS-V} & \multirow{3}{*}{Codec-A} & \multirow{3}{*}{BR} & \multirow{3}{*}{DS-A} & Avg.\\ 
 & & & & & & Bitrate \\
 & & & & & & (kbps) \\
 \hline
 \multirow{7}{*}{\shortstack{H.264 \\(libx264)}} & \multirow{2}{*}{32} & \multirow{2}{*}{4} & \multirow{14}{*}{\shortstack{AAC \\(libfdk\_aac)}} & 5 & 1 & 17.5\\  
  & & & & 10 & 1 & 22.5\\ 
  \cline{2-3} \cline{5-7}
  & \multirow{2}{*}{30} & \multirow{2}{*}{2} & & 5 & 1 & 55.1 \\  
  & & &  & 10 & 1 & 60.1 \\  
    \cline{2-3} \cline{5-7}
  & \multirow{2}{*}{28} & \multirow{2}{*}{2} & & 5 & 1 & 79.8 \\  
  & & & & 10 & 1 & 84.8 \\  
    \cline{2-3} \cline{5-7}
  & \multirow{2}{*}{26} & \multirow{2}{*}{2} & & 5 & 1 & 124.1\\  
  & & & & 10 & 1 & 129.1 \\  
    \cline{2-3} \cline{5-7}
    \cline{1-3}
  \multirow{7}{*}{\shortstack{AV1 \\(libaom-av1)}} & \multirow{2}{*}{63} & \multirow{2}{*}{2} & & 5 & 1 & 13.8\\  
  & & & & 10 & 1 & 18.8\\ 
      \cline{2-3} \cline{5-7}
  & \multirow{2}{*}{63} & \multirow{2}{*}{1} & & 5 & 1 & 16.0\\  
  & & & & 10 & 1 & 21.0\\  
      \cline{2-3} \cline{5-7}
  & \multirow{2}{*}{60} & \multirow{2}{*}{2} & & 5 & 1 & 20.3\\  
  & & & & 10 & 1 & 25.3\\  
 \hline
\end{tabular}
\end{center}
\label{table_codec_params}
\end{table}

\section{Exemplary Encodes} \label{exemplary_encodes}
\begin{figure*}
\centering
\includegraphics[width=1\textwidth]{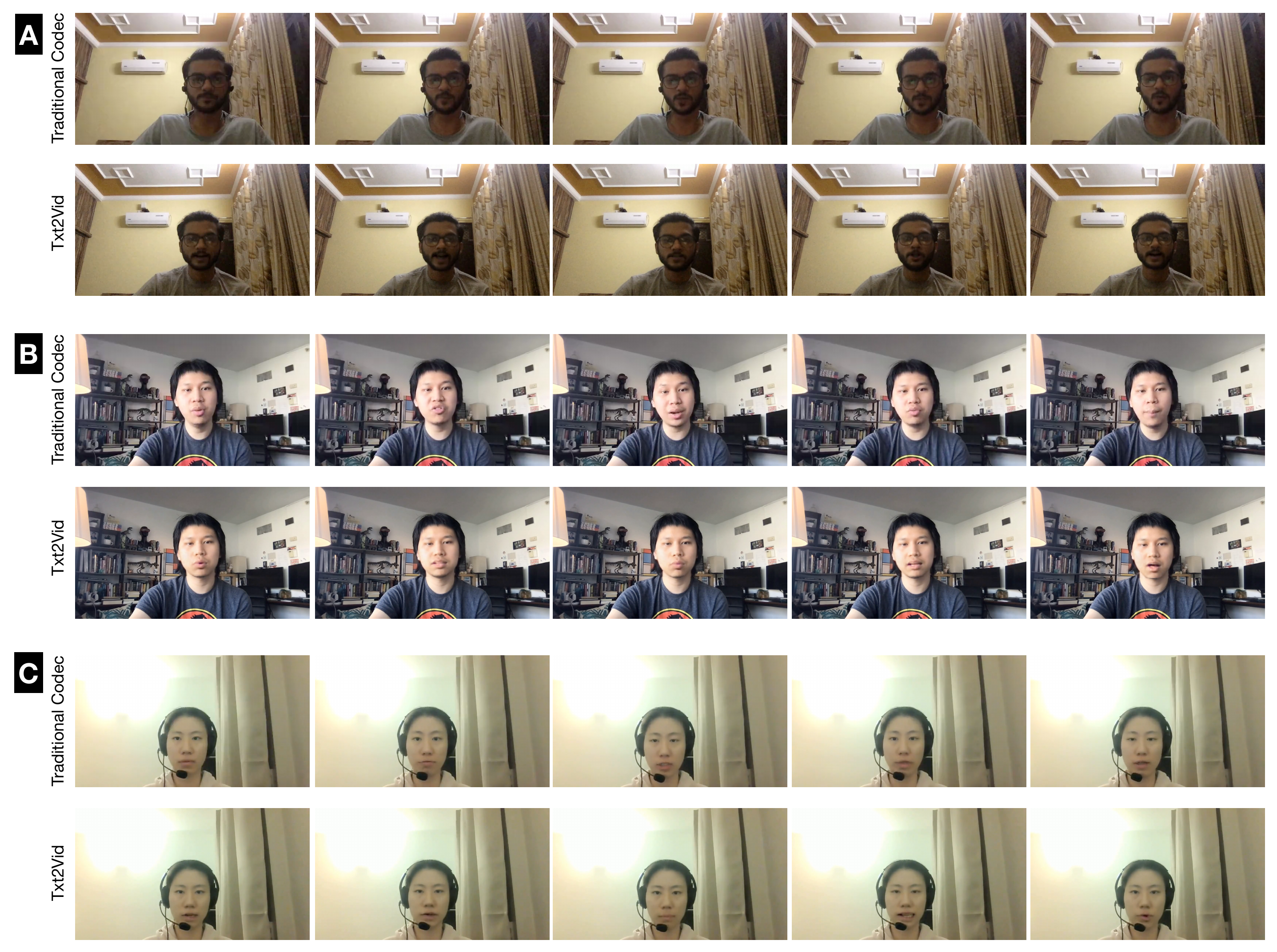}
\caption{Exemplary Encodes. Multiple frames belonging to similar quality-of-experience (QoE) encoded videos generated using traditional audio-video codecs (H.264 + AAC), and  Txt2Vid for three different contents are shown in the top and bottom rows respectively (A: Content 4, B: Content 3, C: Content 6). The videos corresponding to these contents is available on Google Drive (link in Section \ref{sec_intro}).}
\label{fig_examplary_encodes}
\end{figure*}
Figure \ref{fig_examplary_encodes} shows frames belonging to few exemplary encodes for different contents in the dataset obtained at similar-QoE.






\end{document}